\documentclass[sn-mathphys,Numbered]{sn-jnl}

\usepackage{graphicx}%
\usepackage{multirow}%
\usepackage{amsmath,amssymb,amsfonts}%
\usepackage{amsthm}%
\usepackage{mathrsfs}%
\usepackage[title]{appendix}%
\usepackage{xcolor}%
\usepackage{textcomp}%
\usepackage{manyfoot}%
\usepackage{booktabs}%
\usepackage{algorithm}%
\usepackage{algorithmicx}%
\usepackage{listings}%

\newtheorem{lemma}{Lemma}
\newtheorem{theorem}{Theorem}

\usepackage{mathtools}  
\mathtoolsset{showonlyrefs}

\newcommand{\namedref}[2]{\hyperref[#2]{#1~\ref*{#2}}}

\newcommand{\equationref}[1]{\hyperref[#1]{Eq~(\ref*{#1})}}
\newcommand{\theoremref}[1]{\hyperref[#1]{Theorem~\ref*{#1}}}
\newcommand{\lemmaref}[1]{\hyperref[#1]{Lemma~\ref*{#1}}}
\newcommand{\noteref}[1]{\hyperref[#1]{note~\ref*{#1}}}
\newcommand{\appendixref}[1]{\hyperref[#1]{Appendix~\ref*{#1}}}
\newcommand{\corollaryref}[1]{\hyperref[#1]{Corollary~\ref*{#1}}}

\DeclareMathOperator*{\polylog}{polylog}

\DeclareMathOperator{\Min}{Min}

\newcommand{\IN}{\mathbb{N}}

\newcommand{\IE}{\mathbb{E}}

\newcommand{\IR}{\mathbb{R}}

\renewcommand{\vec}[1]{\mathbf{#1}}

\newcommand{\Bc}{\vec{c}}
\newcommand{\Br}{\vec{r}}
\newcommand{\Bp}{\vec{p}}
\usepackage{enumitem}

\begin{document}

\title{Reaching Agreement in Competitive Microbial Systems}

\author[1]{\fnm{Victoria} \sur{Andaur}}\email{victoria.andaur@student-cs.fr}

\author[1]{\fnm{Janna} \sur{Burman}}\email{janna.burman@lri.fr}

\author[2]{\fnm{Matthias} \sur{F\"ugger}}\email{mfuegger@lmf.cnrs.fr}

\author[3]{\fnm{Manish} \sur{Kushwaha}}\email{manish.kushwaha@inrae.fr}

\author[1]{\fnm{Bilal} \sur{Manssouri}}\email{bilal.manssouri@student-cs.fr}

\author[2,4]{\fnm{Thomas} \sur{Nowak}}\email{thomas@thomasnowak.net}

\author[5]{\fnm{Joel} \sur{Rybicki}}\email{joel.rybicki@hu-berlin.de}

\affil[1]{\orgname{Universit\'e Paris-Saclay, CNRS},  \orgaddress{\city{Orsay}, \country{France}}}

\affil[2]{\orgname{Université Paris-Saclay, CNRS, ENS Paris-Saclay}, \orgaddress{\city{Gif-sur-Yvette}, \country{France}}}

\affil[3]{\orgname{Université Paris-Saclay, INRAE, AgroParisTech}, \orgaddress{\city{Jouy-en-Josas}, \country{France}}}

\affil[4]{\orgname{Institut Universitaire de France}, \orgaddress{\city{Paris}, \country{France}}}

\affil[5]{\orgname{Humboldt University of Berlin}, \orgaddress{\city{Berlin}, \country{Germany}}}

\abstract{
We study distributed agreement in microbial distributed systems under stochastic population dynamics and competitive interactions. Motivated by recent applications in synthetic biology, we examine how the presence and absence of direct competition among microbial species influences their ability to reach \emph{majority consensus}.
In this problem, two species are designated as input species, and the goal is to guarantee that eventually only the input species which had the highest initial count prevails.

We show that direct competition dynamics reach majority consensus with high probability even when the initial gap between the species is small, i.e., $\Omega(\sqrt{n\log n})$, where~$n$ is the initial population size. In contrast, we show that absence of direct competition is not robust: solving majority consensus with \emph{constant} probability requires a large initial gap of $\Omega(n)$. To corroborate our analytical results, we use simulations to show that these consensus dynamics occur within practical biological time scales.
}

\maketitle

\section{Introduction}

Competition between reproducing species is ubiquitously observed
  in biological systems and has been extensively studied; see, e.g., \cite{granato2019evolution} for a survey on competition among bacteria.
In the past few decades, the discipline of synthetic biology,
  an area residing at the intersection of biology and engineering,
  has come up with methods to engineer bacteria.
Early success stories in the area showed how to build synthetic genetic toggle switches~\cite{Gal00} and biological oscillators~\cite{Sal08,Nal08}. Recently, the area has rapidly shifted towards distributed computing~\cite{purnick2009second,regot2011distributed,macia2012distributed}: instead of engineering complex metabolic pathways within a single cell, the biological circuit is distributed across a consortium of multiple bacterial strains or species.

Naturally, the decentralization and the engineering of new behavior in bacteria come with new design challenges: computation across different parts of the biological circuit has to be coordinated somehow. While distributed computing theory has dealt with these types of issues for decades, most existing models of distributed computation do not account for the unique features exhibited by microbial systems. For example, microbial communities are subject to stochastic population dynamics, individual bacterial cells reproduce and die at a fast pace, environmental conditions such as available resources rapidly change, and species exhibit complex ecological interactions~\cite{ghoul2016ecology}. 

In this work, we consider distributed computation and coordination in microbiological systems governed by stochastic population dynamics and ecological interactions.
We focus on competitive interactions between multiple species and we examine how the \emph{principle of competitive exclusion} studied in the context of the evolution of biological systems \cite{Hardin60} can be utilized to efficiently solve \emph{majority consensus}---a fundamental task in distributed computing---in microbial populations, be they natural or synthetic, engineered.

In the (binary) consensus problem, the system is initialized with two input values, and the goal is to reach an output configuration where all participants agree on a single value.
In the majority problem, the goal is to output the value that had the majority support initially.

\paragraph{Competitive interactions}
Bacterial communities exhibit a wide range of competitive interactions~\cite{ghoul2016ecology}. 
In order to outcompete other species, bacteria have evolved various types of methods of competition in the course of a biological arms race~\cite{granato2019evolution}.
In this work, we study the presence and absence of \emph{direct or interference competition}. In this case, the individuals use direct means to hinder or fight against competing species. Bacteria may secrete molecules that inhibit the performance of competing species, but they also employ a diverse selection of more lethal methods for interference competition. These range from mechanical weapons that can be used to puncture cell membranes of nearby competitors to using diffusive molecular weapons, such as toxins and antibiotics, that wipe out other species~\cite{granato2019evolution}. In some environments, e.g., in a mammalian gut, some bacterial species are even known to use elaborate schemes to trigger host immune responses to attack against competing strains.

\paragraph{Contributions}
We investigate distributed models of microbiological computation that capture the effects of stochastic population dynamics and competitive interactions. Formally, the population dynamics and protocols are expressed in the \emph{biochemical reaction network} (BCRN) model \cite{briat2016antithetic}, which generalizes \emph{chemical reaction networks} (CRN) \cite{doi:10.1021/j100540a008}.
We analyze population dynamics in BCRNs that allow us to model reactions that do not follow the mass action law,
  and thus better capture complex microbial growth behavior observed in experiments~\cite{Monod1949:growth}.
We emphasize that these types of biological dynamics are not captured by the standard population protocol~\cite{aspnes2009introduction} and CRNs studied in the distributed computing literature.

As our main technical contribution, we extend the coupling technique from~\cite{CFHKNS21:growth} to handle resource-limited models through a \emph{lower-bounding chain} construction, and we adapt the probability analysis to non-self-destructive competition using martingale methods. This allows us to bound the time to competitive exclusion in resource-dependent multi-species models via a simple, discrete-time birth-death process.
We apply this technique to analytically bound the time to reach competitive exclusion.
We model direct competition, where competing species fight each other: on every interaction consisting of two individuals of different species, one of them is killed. 
We show that in this model competitive exclusion occurs and, if the initial discrepancy  between the two species is $\Omega(\sqrt{n\log n})$, then a species with a maximum initial count outcompetes the other species with high probability.
A similar model was studied recently by Cho, Függer, Hopper, Kushwaha, Nowak, and Soubeyran~\cite{CFHKNS21:growth} who investigated approximate majority consensus with direct competitive interactions under birth dynamics.
However, they considered the less biologically realistic setting of self-destructive competition, and our coupling technique allows us to analyze a larger class of biological systems, including  models that incorporate (multiple) resources.

We complete the analytic results with simulations. We study a proof-of-concept hypothetical protocol in engineered bacteria that uses direct competition and, with realistic biological parameters, quickly reaches consensus, demonstrating its practical usability as a tool in synthetic biology.

\paragraph{Outline}
We first review related work in Section~\ref{sec:related}.
In Section~\ref{sec:model}, we discuss biochemical reaction networks (BCRNs) as a basic modeling framework for microbial population dynamics. 
We introduce a new protocol in this formalism, called the \emph{mutual annihilation protocol}, which serves as our main example of microbial dynamics with interference competition between two species.

In Section~\ref{sec:analysis} we first establish an upper bound on the number of steps by coupling the protocol with the discrete-time jump chain. Using this coupling, we show our main result:
  the mutual annihilation protocol achieves majority consensus
  with high probability when the initial gap between the two competing
  cell counts is large enough.
  
In Section~\ref{sec:indirect}, we prove that indirect competition alone is not sufficient to achieve majority consensus with high probability in many cases.
Specifically, in symmetric systems without in-flow of new resources, the probability of reaching majority consensus is equal to the initial relative population of the majority species.

Finally, in Section~\ref{sec:simulations} we provide our simulation case study.

\section{Related Work}\label{sec:related}

We briefly overview related work on majority consensus in molecular and biological models of computation. In particular, we focus on  population protocols and chemical reaction networks.

\subsection{Population Protocols}

Population protocols were designed to study the computational power of an ensemble of indistinguishable agents that interact in an unpredictable manner~\cite{DBLP:journals/dc/AngluinADFP06,DBLP:journals/dc/AngluinAER07}.
In this model, individuals are represented by finite state automata
that communicate via pairwise interactions: in every step, two individuals are chosen to interact. During this interaction, they read each other's states and update their own local states according to a predefined state transition function.

Recently, several papers have investigated fundamental space-time complexity trade-offs in the population protocol model. Space complexity is given by maximum number of local states used by the protocol. Time complexity in this model is studied under a stochastic scheduler that picks two individuals to interact uniformly at random. The (parallel) time complexity is given by the expected number of steps to reach a stable configuration divided by the total population size $n$. 
A long series of papers have investigated how fast majority can be computed by protocols that use a given number of states~\cite{draief2012-convergence,BEFKKR18,DBLP:conf/podc/NunKKP20,doty2020stable}. Many of the recent results and techniques in this area are surveyed by Els\"asser and Radzik~\cite{DBLP:journals/eatcs/ElsasserR18} and Alistarh and Gelashvili~\cite{DBLP:journals/sigact/AlistarhG18}.

The main difference to our setting is that in the stochastic population protocol model the total population size $n$ is assumed to be static and all pairwise interactions occur at the same rate. In contrast, we consider systems, where the population sizes fluctuate and interaction rates may depend on the current configuration of the system. This allows us to more accurately model biological dynamics occurring in microbial populations.

\paragraph{Approximate vs.\ exact majority}
The majority problem has been studied from two perspectives by investigating protocols that compute either \emph{approximate} or \emph{exact} majority.
Both types of protocols typically guarantee consensus, but approximate majority protocols give only probabilistic guarantees on stabilization to the majority opinion, provided that the initial discrepancy between the two initial opinions is sufficiently large. 
In contrast, exact majority protocols guarantee that majority consensus is always eventually reached.

\paragraph{Protocols for approximate majority}
Angluin, Aspnes, and Eisenstat~\cite{angluin2008simple}
analyzed a simple 3-state protocol for solving \emph{approximate} majority in the clique in $O(\log n)$ time provided that the discrepancy between majority and minority value is $\omega(\sqrt{n} \log n)$. Despite its simplicity, the protocol is challenging to analyze. 
Later, Mertzios, Nikoletseas, Raptopoulos, and Spirakis~\cite{mertzios2017determining} studied protocols for approximate agreement when the interactions are restricted to occur on general interaction graphs. In a recent work, Alistarh, Töpfer, and Uznański~\cite{alistarh2021robust} study a generalized approximate majority problem where both initial opinions may have small counts (independent of $n$). They present logarithmic space and time protocols that are self-stabilizing and withstand spurious faulty reactions.
Our results do not directly carry over to population protocols and CRNs: these do not necessarily have single-cell bounded growth for an input/output species.

\paragraph{Protocols for exact majority}
Besides approximate majority, also \emph{exact} majority has been considered in the population protocol setting: regardless of the initial gap, the final outputs should stabilize to the initial majority value. 
Draief and Vojnovi\'c gave a simple 4-state protocol that solves exact majority in any connected interaction graph~\cite{draief2012-convergence}. 
In the clique, the protocol stabilizes in $O(n \log n)$ parallel time.

Alistarh, Gelashvili, and Vojnović~\cite{alistarh2015-fast} gave the first protocols with expected convergence time of the order $\polylog n$ using $\polylog n$ states. 
This result has been gradually improved~\cite{alistarh2017time,BEFKKR18,DBLP:conf/podc/NunKKP20}, and recently, Doty, Eftekhari, and Severson~\cite{doty2020stable} gave protocols that solve majority in $O(\log n)$ time using $O(\log n)$ states.

\subsection{Other Models of Molecular and Biological Dynamics}

Czyzowicz, Gasiniec, Kosowski, Kranakis, Spirakis, and Uznański~\cite{czyzowicz2015convergence} investigated the convergence of discrete Lotka--Volterra dynamics in the population protocol model. These dynamics can be used to model, for example, predator-prey type dynamics. However, they operate in the population protocol model, so while the proportions of the different species can change, the total population size is static throughout.

Condon, Hajiaghayi, Kirkpatrick, and Maňuch~\cite{condon19majority} investigated approximate agreement using multimolecular reactions in the chemical reaction network model. They show that if the gap is $\Omega(\sqrt{n \log n})$, then the majority opinion will with high probability win. However, their work does not consider biological population dynamics.

Closely related to our work is~\cite{CFHKNS21:growth}, who 
showed how to use competitive interactions to solve approximate agreement in a two-species bacterial model in the presence of a gap of $\Omega(\sqrt{n \log n})$. 
Their model assumes unbounded exponential growth without resource limitations, and analyzes self-destructive competition ($A + B \to \emptyset$) where both species die upon interaction. They couple the two-species process to a single-species birth-death M-chain, and exploit the property that annihilation preserves the gap $A - B$ to analyze consensus probability via Yule processes and beta distributions.

We extend their work in two ways. First, we incorporate resource-consumer dynamics through a \emph{lower-bounding chain} construction that abstracts away resource dependencies while preserving worst-case competitive behavior, enabling the coupling approach to be applied to resource-limited models. Second, we analyze non-self-destructive competition ($A + B \to A$ or $B$) where annihilation changes the gap, requiring martingale concentration bounds (Azuma's inequality) rather than beta-distribution arguments.

\section{Preliminaries and Model}
\label{sec:model}

We start with some basic notation. We write $\IN = \{0,1,\dots\}$ for the set of non-negative integers. For functions $f,g : X \to \IR$, we write $f \preceq g$ if $f(x) \leq g(x)$ for all $x \in X$. For two nonempty sets $A$ and $B$, we write $A^B$ to denote the set of functions from $B$ to $A$. When convenient, we treat $\vec x \in A^B$ as a vector with $\lvert B\rvert$ elements.
For a predicate that depends on a parameter~$n$ we say the predicate holds with high
  probability, if there exists a $C> 0$ such that the predicate holds with probability at least $1-n^{-C}$.

The analysis of \emph{single-species discrete-time birth-death chains}
  plays a central role in this work.
Following standard notation, let $p,q \colon \mathbb{N} \to [0,1]$ such that $p(n) + q(n) \le 1$. The birth-death chain defined by $p$ and $q$ is the discrete-time Markov chain $\left( M(k) \right)_{k \ge 0}$ on the state space $\IN$, where in each step the chain goes from state $n$ to $n+1$ with \emph{birth probability} $p(n)$, to state $n-1$ with \emph{death probability} $q(n)$ and stays at state $n$ with \emph{holding probability} $1-p(n)-q(n)$.
We say the chain makes a birth, death, or holding step from state $n$ to the successor state.
A state~$n$ is \emph{absorbing} if $p(n)=q(n)=0$.
We assume throughout that $p(n)>0$ and $q(n)>0$ for all $n >0$ and $p(0)=q(0)=0$ so that state 0 is the unique absorbing state.

\subsection{Biochemical Reaction Networks}

We describe the dynamics of microbial species communities using a generalization of the chemical reaction network (CRN) model, the so-called 
biochemical reaction networks (BCRNs) \cite{briat2016antithetic}.
By contrast to CRNs, BCRNs allow us to model reactions that do not necessarily follow the law of mass action.
A \emph{biochemical reaction network} is a tuple $\mathcal{B} = (S,\mathcal{R},v)$, where $S$ is the set of species,  $\mathcal{R}$ is a set of \emph{reactions} that may occur between the different entity types, and~$v$ is the volume of the system. For simplicity, we assume unit volume $v=1$, unless otherwise specified.

A \emph{configuration} is a vector $\vec c \in \IN^S$, where $c_i$ denotes the count of entities of type $i \in S$ in the configuration $\vec c$. A \emph{reaction} is a triple $(\vec r, \vec p, \alpha)$, where $\vec r \in \mathbb{N}^S$ describes the reactants and $\vec p \in \mathbb{N}^S$ the products of the reaction. The entity type $i$ is said to be a \emph{reactant} if $r_i > 0$ and a \emph{product} if $p_i > 0$. Note that an entity can be both a reactant and a product in the same reaction.
A reaction is \emph{applicable} to configuration~$\vec x$ if $r_i \leq x_i$ for each $i \in S$, that is, the reactants are present in the configuration~$\vec x$.  The map $\alpha \colon \IN^S \to \IR_{\geq0}$ is the \emph{propensity function} of the reaction and $\alpha(\vec c)$ is the propensity of the reaction in configuration $\vec c$. If the reaction is not applicable to $\vec c$, then $\alpha(\vec c)=0$. Given a normalized (dimensionless) volume $v=1$, the \emph{rate} of a reaction is equal to its propensity. We often use the common notation
\[
\sum_{i \in S}{r_i}\cdot i
\xrightarrow{\makebox[0.6cm]{$\alpha$}}
\sum_{i \in S}{p_i} \cdot i
\]
to describe a reaction $(\vec r, \vec p, \alpha)$.
  We assume that there are two kinds of propensity functions: \emph{growth reactions} and \emph{mass-action reactions}.
They are specified as follows:
\begin{enumerate}
\item \emph{Growth reactions:}
Let $\overline{{\mathscr T}} = {\mathscr S} \setminus {\mathscr T}$ be the species that are not cell types.
A growth reaction for cell type $T \in {\mathscr T}$ is of the
  form
\begin{equation*}
T + \sum_{s \in \overline{{\mathscr T}}}\Br(s)
\xrightarrow{\makebox[0.6cm]{$\alpha$}}
2T + \sum_{s \in \overline{{\mathscr T}}}\Bp(s)\enspace.
\end{equation*}
Intuitively the reactants from $\overline{\mathscr T}$ account for resources and waste products in the growth medium that are (potentially) consumed
  and that determine the cell's growth rate.
Assuming that cells grow independently of each other, we have that
\begin{equation*}
\alpha(\Bc) = \Bc(T) \cdot \gamma(\Bc) \enspace,
\end{equation*}
where $\gamma(\Bc)$ is the \emph{individual cell growth rate}.
We further assume that $\gamma(\Bc)$ is bounded by a maximal growth rate $\Gamma$.
This is motivated by the observation that cells do not increase their growth rate arbitrarily
  with the number of available resources in the medium.
  
\item \emph{Mass-action reactions:}
For reactions other than growth reactions, we assume the classical mass-action kinetics to hold.
Their propensity function $\alpha$ is given as
\begin{equation*}
\alpha(\Bc) = \frac{\xi}{v^{o-1}} \prod_{s\in {\mathscr S}} \binom{\Bc(s)}{\Br(s)}\enspace,
\end{equation*}
where $o$ is the order of the reaction,
$\binom{\Bc(s)}{\Br(s)}$ denotes the binomial coefficient of~$\Bc(s)$ and~$\Br(s)$,
and $\xi \geq 0$ is the \emph{rate constant} of the reaction.
We follow the convention that the binomial coefficient is~$1$ if $\Br(s)=0$, and
it is~$0$ if $\Br(s) > \Bc(s)$.
By slight abuse of notation we write
$A + \dots \xrightarrow{\makebox[0.6cm]{$\xi$}} B + \dots$
instead of
$A + \dots \xrightarrow{\makebox[0.6cm]{$\alpha$}} B + \dots$
for mass-action reactions.

\end{enumerate}

\paragraph{Stochastic dynamics}
Let $\vec X = (\vec X(t))_{t \ge 0}$ be the continuous-time Markov process on the set $\mathbb{N}^S$ of the configurations of the BCRN with the transition rate matrix $Q$ determined as follows:
Given that the BCRN is in configuration~$\vec x$ at time $t \ge 0$, the new configuration after an applicable reaction $(\Br, \Bp, \alpha)$ is equal to  $\vec x' = \vec x - \vec r + \vec p $ and it transitions to this configuration with the rate of the reaction. For any configuration $\vec x \in \mathbb{N}^S$ we set
\[
Q(\vec x) = \sum_{\vec y \neq \vec x} Q(\vec x,\vec y)\enspace.
\]
For $\vec x \neq \vec y$, the transition probability $P(\vec x,\vec y)$ of moving from configuration $\vec x$ to $\vec y$ is given by
$P(\vec x,\vec y) = Q(\vec x,\vec y)/Q(\vec x)$,  which defines the transition probability matrix of the discrete-time \emph{jump chain} $\vec Y = (\vec Y(k))_{k \ge 0}$ of the process $\vec X$.

\paragraph{Majority consensus}
A configuration $\vec x$ has reached \emph{consensus} if all but (at most) one cell type ($A$ or $B$) has gone extinct.
We define the \emph{consensus time} as the random variable
\[
C(\vec x) =  \inf \{ t : \vec X(0) = \vec x
\textrm{ and }
\vec X(t) \textrm{ has reached consensus at time } t \}
\enspace.
\]
We say that  \emph{majority consensus} was reached in a
  trajectory with initial configuration~$\vec x$ if
  (i) time $T = C(\vec x)$ is finite and (ii) the cell type that is not extinct at time $T$ had a maximal initial count.

For an initial configuration $\vec x$, we define the \emph{majority consensus probability} $\rho(\vec x)$ to be the probability that majority consensus is reached from initial configuration $\vec x$.
We are particularly interested in bounding the probability as a function of the initial discrepancy of the cell counts of $A$ and $B$ in the initial configuration.

\subsection{Multiple-Resource Protocols with Cell Types}

Cells replicate depending on the resources within the medium
  with a certain rate.
Typically such resource-dependent behavior is modeled with
  cell growth rates that depend on one or several limiting resources~\cite{Monod1949:growth}.
In the following we assume growth dependent on multiple resource types.
This is biologically plausible for \emph{Escherichia coli} grown under lab conditions that were observed to feed on two main nutrient types, a resource enabling initial fast growth (called $R_1$ in the following) and one that leads to slow growth (denoted by $R_2$), showing so-called diauxic growth behavior \cite{sezonov2007escherichia}.
Our results, however, generalize to more than two resources.

In our model such a system is a BCRN whose species are partitioned into
  proliferating cells with birth and death reactions, and a set
  of~$k$ passive resources.
For that purpose, we define:
Let $\gamma_1, \dots, \gamma_k$ be propensity functions.
A \emph{multiple-resource protocol with cell types~$\mathscr{T}$}
  is a BCRN with set of species ${\mathscr S} = \{R_1, \dots, R_k\} \cup {\mathscr T}$,
  cell types $\mathscr T$, and for every cell type $X\in\mathscr{T}$, the reactions
\begin{align*}
    X + R_i \xrightarrow{\makebox[0.6cm]{$\gamma_i$}} 2X
    \enspace.
\end{align*}

\medskip

In this work we analyze a specific multiple-resource protocol with two competing cell types, the \emph{mutual annihilation protocol}.
The protocol is exemplary for interference competition between two bacteria types $A$ and $B$.
While this protocol can be seen as an instance of naturally occurring direct attacks between $A$ and $B$, such a protocol also lends itself to be implemented in a system of synthetic bacteria.
For a proof-of-principle design for engineered direct competition, assume the
two types of \emph{E. coli}, $A$ and $B$, are engineered to transfer small circular DNA in the form of plasmids upon contact into the receiving cell, a process known as bacterial conjugation \cite{thomas2005mechanisms}.
Further, we assume that the plasmids are designed such that their presence leads to the death of the receiving \emph{E. coli} if it is of a different type.
We call this protocol the mutual annihilation protocol.

Formally, the mutual annihilation protocol is a multiple-resource protocol with cell types~${\mathscr T} = \{A,B\}$
  with the additional mass-action reactions
\begin{align}
    & A+B \xrightarrow{\makebox[0.6cm]{$\alpha$}} A \\
    & A+B \xrightarrow{\makebox[0.6cm]{$\alpha$}} B\enspace,
\end{align}
where $\alpha > 0$ is the annihilation rate constant.
This parameter is used to model the effects of direct interference competition between two species. In particular this can be used to model, for example, mechanical attacks that puncture the cell membranes of opposing species, but also mechanisms relying on bacterial conjugation.

Intuitively this protocol should strongly amplify any
  difference between the initial concentrations
  of $A$ and $B$ until finally majority consensus is achieved.
The rest of the paper is devoted to showing that this is
  indeed the case.

\section{Analysis of the Mutual Annihilation Protocol}
\label{sec:analysis}

In this section, we analyze the mutual annihilation protocol. First, we bound the number of steps until consensus (competitive exclusion). Afterwards, we show that if the initial gap is large enough, then majority consensus is reached with high probability.

\subsection{Reduction to Single-Species Process}
\label{sec:stepsbound}

In this section, we upper-bound the number of 
  steps that a reaction is applied in the mutual
  annihilation protocol, resulting in a change
  of species counts,  until consensus is achieved.
This step upper bound is the main ingredient for our
  probability bound to achieve majority consensus.
  
Towards this goal we introduce three Markov chains
  that gradually abstract the behavior of the
  mutual annihilation protocol while maintaining
  central stochastic properties:
  the $\mathscr S$-chain, the lower-bounding $\mathscr S$-chain,
  and the $M$-chain.

\subsubsection{The $\mathscr S$-Chain of the Mutual Annihilation Protocol}

Let $\mathscr S$ be a set of species of a BCRN.
Consider the discrete-time jump chain $\left( \vec S(k) \right)_{k \ge 0}$ on the state space of the BCRN's configurations $\mathscr{C} = \IN^{\mathscr{S}}$ such that $\vec S_s(k)$ is a random variable denoting the number of individuals of species $s \in \mathscr{S}$.
A configuration $\vec c \in \mathscr{C}$ gives the quantity of each species so that $c_s$ is the number of individuals of species~$s$. 
We call any such process an \emph{$\mathscr S$-chain}.

The mutual annihilation protocol includes the cell types~$A$ and~$B$ as well as the resource species~$R_i$.
Its stochastic behavior is fully described via
  a corresponding $\mathscr S$-chain with species $\mathscr S = \{A,B,R_1,\dots,R_k\}$.

\subsubsection{The Lower-Bounding $\mathscr S$-Chain}\label{sec:lower:bounding:s:chain}

We next define an abstraction of an $\mathscr S$-chain, for
  the particular case of the $\mathscr S$-chain of the
  mutual annihilation protocol.
Let~$\Gamma$ be the uniform upper bound on the sum of the birth rates~$\gamma_i(r_i)$ of cells $A$ and $B$, respectively, which exists by the assumption on BCRN growth reactions.

Given an initial state~$(a,b,r_1,\dots,r_k)$ of the $\mathscr S$-chain of the mutual annihilation protocol with $A(0) = a > b = B(0)$, we define the \emph{lower-bounding $\mathscr S$-chain} as a two-species chain with initial state~$(a,b)$.
In this chain, the majority species~$A$ has birth rate~$0$ and the minority species~$B$ has birth rate~$\Gamma$.
There are no resource species in the lower-bounding chain.

In a state $(a,b)$ of the lower-bounding $\mathscr S$-chain, there are three possible transitions:
birth of species~$B$ with rate $b\Gamma$,
annihilation of~$A$ with rate $ab\alpha$,
and
annihilation of~$B$ with rate $ab\alpha$.
The probability of a birth of species~$B$ is thus equal to $b\Gamma/(b\Gamma + 2ab\alpha)$.
The annihilation of species~$A$ and of species~$B$ each have probability equal to $ab\alpha/(b\Gamma + 2ab\alpha)$.

It is straightforward to show that the time to reach majority consensus in the lower-bounding $\mathscr S$-chain is an upper bound on the time to reach majority consensus in the $\mathscr S$-chain of the mutual annihilation protocol.

\subsubsection{The $M$-Chain}

Towards the goal of abstracting the lower-bounding $\mathscr S$-chain, we introduce the
  $M$-chain which tracks the population size of the cell type
  that is currently the minimum among the cell types.
  
Formally, an \emph{$M$-chain} is a discrete-time birth-death Markov process
$\left( M(k) \right)_{k \ge 0}$ on the state space $\IN$
  with birth probability function $p' : \IN \to [0,1]$
  and death probability function $q' : \IN \to [0,1]$,
  where $p'(m) + q'(m) \leq 1$ for all $m \in \IN$:
a population of size $m \geq 0$ increases by one in the next step with
  probability $p'(m)$ and decreases in the next step with probability~$q'(m)$.

Consider an $\mathscr S$-chain with the two cell
  types $A$ and $B$.
For such an $\mathscr S$-chain define the sequence of random variables
\[
\Min(k) = \min \left\{ \vec{S}_A(k), \vec{S}_B(k) \right\}
\]
and set $p,q \colon \mathscr{C} \to [0,1]$ as follows:
\begin{align*}
p(\vec c) &= \Pr[\Min(k+1) = \Min(k)+1 \mid \vec{S}(k) = \vec c] \\
q(\vec c) &= \Pr[\Min(k+1) = \Min(k)-1 \mid \vec{S}(k) = \vec c]\enspace.
\end{align*}
That is, $p(\vec c)$ gives the probability of the process transitioning from state $\vec c$ to a state $\vec c'$, where the minimum of $c_A$ and $c_B$ increases by one in the next step.
Analogously, $q(\vec c)$ gives the probability that the minimum decreases by one during the next transition.
Note that the probability of $\Min(k)$ increasing or decreasing may depend on the entire configuration $\vec S(k)$.

We say an $M$-chain \emph{dominates} the $\mathscr S$-chain if
  for all $\vec c \in \IN^{\mathscr{S}}$
  with $m = \min\{\vec{c}_A, \vec{c}_B\}$,
  functions $p', q'$ satisfy
\begin{align}
  p(\vec c) &\le p'(m) \label{eq:pq-A1}\\
  q(\vec c) &\ge q'( m ) \label{eq:pq-A2}\\
  p(\vec c) &\le 1-q'( m+1 )\enspace.\label{eq:pq-A3}
\end{align}

\subsubsection{Coupling the $\mathscr S$-Chain to the $M$-Chain}

In the following we construct a coupling $(\widehat{\vec S}, \widehat{M})$ of the 
  $\mathscr S$-chain and a dominating $M$-chain.
Analogously to the variable $\Min$, we define the variable
\[
\widehat{\Min}(k) =
\min \left\{ \widehat{\vec S}_{A}(k), \widehat{\vec S}_{B}(k) \right\}
\]
for all $k \ge 0$ on the coupling.
We will next define the discrete time processes $\widehat{\vec S}$ and $\widehat{M}$ inductively.

\smallskip

\noindent {\bf Initially:} Set $\widehat{\vec S}(0) = \vec{S}(0)$ and
$\widehat{M}(0) = M(0)$. 

\smallskip

\noindent {\bf Step:}
Let $(\xi(k))_{k \in \IN}$ be a sequence of i.i.d.\ random values sampled uniformly from the unit interval $[0,1)$.
Let $k \in \IN$.
Assuming $\left( \widehat{\vec S}(k),\widehat{M}(k) \right) = (\vec{c},m)$, we set
   $\left( \widehat{\vec S}(k+1),\widehat{M}(k+1) \right)$ as follows:
\begin{description}
    \item[Minimum increases.] If $\xi(k) \in [0, p(\vec c))$, then sample $\widehat{\vec  S}(k+1)$ according to the conditional distribution
    \[
     \mu(\vec{c}') = \Pr\left[ \widehat{\vec  S}(k+1) = \vec{c}' \mid \widehat{\vec  S}(k) = \vec{c} \,\textrm{ and }\, \widehat{\Min}(k+1) = \widehat{\Min}(k) + 1 \right]\enspace.
    \]
    If $\xi(k) \in [0, p'(m))$, then set $\widehat{M}(k+1) = \widehat{M}(k) + 1$.
    
    \item[Minimum decreases.] If $\xi(k) \in [1-q(\vec c), 1)$, then sample  $\widehat{\vec  S}(k+1)$ according to the conditional distribution
    \[
    \mu(\vec{c}') = \Pr\left[ \widehat{\vec  S}(k+1) = \vec{c}' \mid \widehat{\vec  S}(k) = \vec c \,\textrm{ and }\,  \widehat{\Min}(k+1) = \widehat{\Min}(k) - 1 \right]\enspace.
    \]
   If $\xi(k) \in [1-q'(m), 1)$, then set  $\widehat{M}(k+1) = \widehat{M}(k) - 1$.
    
    \item[Minimum does not change.] Otherwise, sample $\widehat{\vec  S}(k+1)$ according to the conditional distribution
    \[
    \mu(\vec{c}') = \Pr\left[ \widehat{\vec  S}(k+1) = \vec{c}' \mid \widehat{\vec  S}(k) = \vec{c} \,\textrm{ and }\, \widehat{\Min}(k+1) = \widehat{\Min}(k) \right]
    \]
    and set $\widehat{M}(k+1) = \widehat{M}(k)$.
\end{description} 
Observe that by construction, the marginal distributions of $\widehat{\vec S}(k)$
  and $\widehat{M}(k)$ equal the distributions of $\vec S(k)$ and $M(k)$
  for all $k \in \IN$.

We will next show that $\widehat{\Min} \preceq \widehat{M}$ in the coupled process
  under certain dominance conditions of the transition probabilities in the
  original $\mathscr S$-chain and a dominating $M$-chain.

\begin{lemma}
\label{lem:coupling-M-chain}
Given an $\mathscr S$-chain and a dominating $M$-chain,
$\Min(0) \le M(0)$ implies $\widehat{\Min} \preceq \widehat{M}$.
\end{lemma}
\begin{proof}
We show by induction that $\widehat{\Min}(k) \le \widehat{M}(k)$ for all $k \in \IN$.
The base case $k=0$ is vacuous. 
For the inductive step, suppose $\widehat{\Min}(k) \le \widehat{M}(k)$ holds for some $k \ge 0$.
Observe that since $|\widehat{\Min}(k+1)-\widehat{\Min}(k)| \le 1$
  and $|\widehat{M}(k+1)-\widehat{M}(k)| \le 1$,
  the claim follows if:
\begin{align}
\widehat{\Min}(k+1) &< \widehat{\Min}(k) \enspace\text{or} \label{eq:C2}\\
\widehat{M}(k+1) &> \widehat{M}(k)\enspace. \label{eq:C3}
\end{align}
Let $\vec c = \widehat{\vec S}(k)$ and $m = \min \{ \vec{c}_A, \vec{c}_B \}$.
We distinguish two cases:
\begin{enumerate}
    \item Suppose $\widehat{\Min}(k) = \widehat{M}(k) = m$.
    To show that $\widehat{\Min}(k+1) \le \widehat{M}(k+1)$, we consider the following subcases:
    \begin{enumerate}
        \item If $\widehat{\Min}(k+1) = m+1$, then $\xi(k) \in [0, p(\vec c)) \subseteq [0,p'(m))$, by Assumption~\eqref{eq:pq-A1} of the lemma.
        Now the update rule of the coupling yields $\widehat{M}(k+1) = m+1 > \widehat{M}(k)$; i.e., \eqref{eq:C3} is fulfilled and the claim follows.
        
        \item  If $\widehat{M}(k+1) = m-1$, then $\xi(k) \in [1-q'(m), 1) \subseteq [1-q(\vec c),1)$, by Assumption~\eqref{eq:pq-A2} of the lemma.
        Thus, the update rule implies $\widehat{\Min}(k+1) = m-1 < \widehat{\Min}(k)$; i.e., \eqref{eq:C2} is fulfilled
        and the claim follows.
        
        \item Otherwise, $\widehat{\Min}(k+1) \leq m$ and $\widehat{M}(k+1) \geq m$; the claim
          follows in this case.
    \end{enumerate}
    Thus, in all three cases the claim follows.
    
    \item Otherwise, by the induction hypothesis, $\widehat{\Min}(k) < \widehat{M}(k)$.
    If $\widehat{M}(k)- \widehat{\Min}(k) > 1$, then the claim follows immediately, since both variables can change by at most one per step and thus no reordering can happen.
    
    Hence, suppose that $\widehat{M}(k) = \widehat{\Min}(k) + 1$ holds.
    The only remaining case is the event where
    $\widehat{\Min}(k+1) = \widehat{\Min}(k)+1 = m+1$ and
    $\widehat{M}(k+1) = \widehat{M}(k) -1 = m$.
    This implies that $\xi(k) \in [0, p(\vec c)) \cap [1-q'(m+1), 1)$.
    Thus, $p(\vec c) > 1 - q'(m+1)$,
    contradicting Assumption~\eqref{eq:pq-A3} of the lemma.
    Therefore, the case where $\widehat{\Min}$ increments and
    $\widehat{M}$ decrements does not occur. \qedhere
\end{enumerate}
\end{proof}

\subsubsection{An $M$-Chain for the Lower-Bounding $\mathscr S$-Chain}

We first define~$p'(m)$ by setting
\begin{equation}
p'(m)
=
\frac{m \Gamma}{m\Gamma + 2m^2\alpha}\enspace.
\end{equation}
The maximum of~$p'$ is achieved for $m = 1$.
This maximum is strictly smaller than~$1$ because $p'(m) < 1$ for all $m\in\IN$.
Call the maximum~$P$.

We then define~$q'(m)$ by setting
\begin{equation}
q'(m)
=
\min\left\{
1-P
\ ,\ 
\frac{m^2\alpha}{m\Gamma + 2m^2\alpha}
\right\}
\enspace.
\end{equation}

We observe that $p'(m) = O(1/m)$ and $q'(m) = \Omega(1)$.
We next show dominance for the lower-bounding chain of the mutual annihilation protocol.

\begin{lemma}
The functions~$p'$ and~$q'$ define an $M$-chain that dominates the lower-bounding chain of the mutual annihilation protocol.
\end{lemma}
\begin{proof}
We first prove Condition~(1).
Assume $m = b < a$.
Using the probabilities calculated in Section~\ref{sec:lower:bounding:s:chain}, we have:
\begin{equation}
\begin{split}
p(a,b)
& =
\frac{b\Gamma}{b\Gamma + 2ab\alpha}
<
\frac{b \Gamma}{b\Gamma + 2b^2\alpha}
=
p'(b) = p'(m)
\end{split}
\end{equation}
If $m = a \leq b$, then $p(a,b)=0$ and the inequality trivially holds.

We next prove Condition~(2).
Assume $m = b < a$.
Using the probabilities calculated in Section~\ref{sec:lower:bounding:s:chain} we have:
\begin{equation}
\begin{split}
q(a,b)
& =
\frac{ab\alpha}{b\Gamma + 2ab\alpha}
=
\frac{b\alpha}{\Gamma + 2b\alpha}
=
\frac{b^2\alpha}{b\Gamma + 2b^2\alpha}
\geq
q'(b) = q'(m)
\end{split}
\end{equation}
If $m = a \leq b$, then:
\begin{equation}
\begin{split}
q(a,b)
& =
\frac{ab\alpha}{b\Gamma + 2ab\alpha}
=
\frac{a\alpha}{\Gamma + 2a\alpha}
=
\frac{a^2\alpha}{a\Gamma + 2a^2\alpha}
\geq
q'(a) = q'(m)
\end{split}
\end{equation}

Lastly, Condition~(3) easily follows from Condition~(1) and the definition of~$q'(m)$ since
\begin{equation}
p(a,b) + q'(m+1) \leq p'(m) + (1-P) \leq P + (1-P) = 1
\end{equation}
where $m = \min\{a,b\}$. 
\end{proof}

\subsubsection{Number of Steps in the $M$-Chain}

We now show that the number of steps until extinction in the $M$-chain is at most linear in the initial population, both in expectation (Lemma~\ref{lem:m:chain:expected}) and with high probability (Lemma~\ref{lem:m:chain:whp}).
Because the $M$-chain dominates the $\mathscr S$-chain, the upper bound also holds for the number of steps until consensus in the lower-bounding $\mathscr S$-chain.

\begin{lemma}\label{lem:m:chain:expected}
The expected number of steps until extinction of any $M$-chain with $p'(m) = O(1/m)$ and $q'(m) = \Omega(1)$
is~$O(M)$ where~$M$ is the initial state.
\end{lemma}
\begin{proof}
Let $C\geq 1$ be a big-oh constant for~$p'(m)$, i.e., $p'(m) \leq C/m$ for all $m \geq 1$.
Let $D>0$ be a big-omega constant for~$q'(m)$, i.e., $q'(m) \geq D$ for all $m\geq 1$.
From known results for discrete-time birth-death processes~\cite{karlin75}, setting $\alpha = C/D$, we get that the expected number of steps until extinction from initial state~$M$ is equal to
\begin{equation}
\begin{split}
\sum_{j=1}^M \sum_{k=j-1}^\infty \frac{p'(j)\cdots p'(k)}{q'(j)\cdots q'(k+1)}
& \leq
\sum_{j=1}^M \sum_{k=j-1}^\infty \frac{C^{k-j+1} (j-1)!}{D^{k-j+2} k!}
=
\frac{1}{D}\sum_{j=1}^M \sum_{k=j-1}^\infty \alpha^{k-j+1} \frac{(j-1)!}{k!}
\\ & \leq
\frac{1}{D}\sum_{j=1}^M \sum_{k=j-1}^\infty \alpha^{k-j+1} \frac{1}{(k-j+1)!}
=
\frac{1}{D}\sum_{j=1}^M \sum_{k=0}^\infty \alpha^{k} \frac{1}{k!}
\\ & =
\frac{1}{D}\sum_{j=1}^M  e^{\alpha}
=
O(M)\,.
\end{split}
\end{equation}
Here, we used the inequality $\frac{(j-1)!}{k!} \leq \frac{1}{(k-j+1)!}$, which is equivalent to $\binom{k}{j-1}\geq 1$.
\end{proof}

\begin{lemma}\label{lem:m:chain:whp}
The number of steps until extinction of
any $M$-chain with $p'(m) = O(1/m)$ and $q'(m) = \Omega(1)$
is~$O(M)$ with probability $1-O(1/\sqrt{M})$ where~$M$ is the initial state.
\end{lemma}
\begin{proof}
We distinguish two phases: the first from states~$M$ to $\Theta(\sqrt{M})$ and the second from $\Theta(\sqrt{M})$ to~$0$.

For the first phase, we start by bounding the number of holding reactions and then the number of birth steps to show that enough death reactions occur.
Let $D>0$ be a big-omega constant for~$q'(m)$, i.e., $q'(m) \geq D$ for all $m\geq 1$.
The probability of an individual step being a holding step is at most $\beta = 1-D < 1$.
Let $C\geq 1$ be a big-oh constant for~$p'(m)$, i.e., $p'(m) \leq C/m$ for all $m \geq 1$.
We pose~$KM$ as an upper bound on the number of steps of the first phase where $K \geq \frac{2}{1-\beta}$.

The expected number of holding steps in the first~$KM$ steps is at most $\mu \leq \beta K M$.
Setting $\delta = \frac{1-\beta}{2\beta}$,
by the Chernoff bound, the probability of having more than $(1+\delta)\beta K M = \frac{1+\beta}{2}K M$ holding steps in the first~$KM$ steps is upper-bounded by
$e^{-\delta^2\beta K M/3} = e^{-\Omega(M)}$.
By the choice of~$K$, the same bound holds for the probability that there are less than~$M$ non-holding steps in the first~$KM$ steps.

The first phase ends when a state $\leq 4C\sqrt{M}$ is reached.
We have $m \geq \sqrt{M}$ and thus $p'(m) \leq C/\sqrt{M}$ in particular in the first phase.
Let~$E$ be the event that a state $\leq 4C\sqrt{M}$ is reached in the first~$M$ non-holding steps.
The event that the number~$b$ of births in the first~$M$ non-holding steps is at most $2C\sqrt{M}$ implies event~$E$. 
Therefore, the inverse event $\neg E$ implies $b > 2C\sqrt{M}$.
By the Chernoff bound, 
the probability of $\neg E$ is bounded by
\begin{equation}
\Pr[\neg E]
\leq
\Pr[b \geq 2C\sqrt{M} ]
=
\Pr[b \geq (1+\delta)\mu ]
\leq
e^{- \delta^2\mu/3}
\end{equation}
where $\mu \leq  C\sqrt{M}$ and $\delta = 1$.
We thus have:
\begin{equation}
\Pr[\neg E]
=
e^{-\Omega\left( \sqrt{M} \right)}
\end{equation}

In the second phase, denote by~$L$ the number of events until extinction.
By Lemma~\ref{lem:m:chain:expected}, the expected value of~$L$ is upper-bounded by $\IE\,L = O(\sqrt{M})$.
By Markov's inequality we thus have:
\begin{equation}
\Pr[L > M]
\leq
\frac{\IE\,L}{M}
=
O(1/\sqrt{M})
\end{equation}

Combining the analyses of both phases shows that extinction happens in the first $(K+1)M$ steps with high probability.
\end{proof}

\subsection{Number of Steps Until Consensus}

We can now prove our step bound for reaching consensus with the mutual annihilation protocol.

\begin{lemma}\label{lem:s:chain:steps}
The number of steps until consensus in the $\mathscr S$-chain is~$O(n)$ with high probability where~$n$ is the total initial population size.
\end{lemma}
\begin{proof}
The number of steps until consensus with a smaller initial gap and the same total population stochastically dominates the number of steps until consensus with a larger initial gap.
It is hence sufficient to prove the lemma for a constant initial gap.
But then $a = \Omega(n)$ and $b = \Omega(n)$ initially.
The corresponding $M$-chain thus has initial state $m=\Theta(n)$.
An invocation of Lemma~\ref{lem:m:chain:whp} concludes the proof.
\end{proof}

\subsection{Probability of Reaching Majority Consensus}
\label{sec:probability_consensus}

Equipped with the step upper bound from Lemma~\ref{lem:s:chain:steps},
we prove our probability bound for majority consensus in this subsection.
The proof is based on Azuma's inequality for sub-martingales.
A real-valued stochastic process $X = \big( X(k) \big)_{k\geq 0}$ is a sub-martingale if $\IE\big[ X(k+1) \mid X(1), \dots, X(k) \big] \geq X(k)$.

\begin{theorem}[Azuma's inequality]\label{thm:azuma}
Let $X = \big( X(k) \big)_{k\geq 0}$ be a sub-martingale such that 
$\lvert X(k) - X(k-1) \rvert \leq c_k$ almost surely.
Then, for all $K\in\IN$ and all $\varepsilon > 0$ we have
\begin{equation}
\Pr[ X(0) - X(K) \geq \varepsilon ]
\leq
e^{-\varepsilon^2/(2\sum_{k=1}^K c_k^2)}
\enspace.
\end{equation}
\end{theorem}
\begin{proof}
Follows from \cite[Theorem~13.4]{MU17}.
\end{proof}

\begin{lemma}\label{lem:submartingale}
Let $A(0) \geq B(0)$ and set $X(k) = A(k) - B(k)$.
As long as $X(k) \geq 0$, the process $(X(k))$ is a sub-martingale.
\end{lemma}
\begin{proof}
We have
\begin{equation}
\begin{split}
\IE[X(k+1) - X(k) \mid X(k)]
& =
\frac{\gamma(\vec S(k))\cdot A(k) - \gamma(\vec S(k)) \cdot B(k)}{Z(\vec S(k))}
=
\frac{\gamma(\vec S(k))\cdot X(k)}{Z(\vec S(k))}
\enspace,
\end{split}
\end{equation}
where $\gamma$ is the individual cell growth rate in state $\vec S(k)$ and
  $Z(\vec S(k))$ is the sum of all propensities in state $\vec S(k)$.
Hence if $X(k)\geq0$ then $\IE[X(k+1) - X(k) \mid X(k)]\geq 0$.
\end{proof}

\begin{theorem}\label{thm:mutual:prob}
If the species~$A$ and~$B$ are symmetric,
there is a constant $D>0$ such that
the mutual annihilation protocol achieves majority consensus with high probability whenever the initial gap is $\geq D\sqrt{n \log n}$ where~$n$ is the total initial population size.
\end{theorem}
\begin{proof}
With high probability, the number~$K$ of steps until consensus is $\leq C n$ by Lemma~\ref{lem:s:chain:steps}.
Set $D = 2\sqrt{C}$.
Without loss of generality, let~$A$ be the initial majority species.
By hypothesis, we have $X(0) = A(0) - B(0) \geq D\sqrt{n\log n}$.
The maximum step size of the process~$(X(k))$ is bounded by $\lvert X(k) - X(k-1) \rvert \leq 1$.
Set~$\varepsilon$ equal to the initial gap, i.e., $\varepsilon = D\sqrt{n \log n}$.
By Lemma~\ref{lem:submartingale}, we can apply the union bound and Azuma's inequality (Theorem~\ref{thm:azuma}) to get:
\begin{equation}
\begin{split}
\Pr[ X(K) < 0 ]
& \leq
\Pr[ \exists k\leq K\colon X(k) < 0]
\\ &
\leq
\sum_{k=1}^K
\Pr[X(k) < 0 \wedge X(k-1) \geq 0 \wedge \dots \wedge X(1) \geq 0 ]
\\ & \leq
K\exp\left( \frac{-\varepsilon^2}{2\sum_{k=1}^K 1} \right)
\leq
\exp\left( \log C + \log n - \frac{D^2n\log n}{2Cn} \right)
\\ &
=
\exp\left( \log C - \left(\frac{D^2}{2C}-1\right)\log n \right)
\leq
\exp\big( \log C - \log n \big)
=
1/n^{\Omega(1)} \enspace .
\qedhere
\end{split}
\end{equation}
\end{proof}

\section{Inefficiency of Indirect Competition}\label{sec:indirect}

In this section, we evaluate the performance of indirect competition, which does not employ any direct interactions between the bacterial species and relies on indirect competition via shared resources only, for majority consensus.
For a species to be able to die, and thus achieve majority consensus, we have to add individual death reactions.
We thus assume the presence of death reactions $A\to \emptyset$ and $B\to\emptyset$ with mass action kinetics.
For simplicity, we will first assume equal birth rates and death rates for both species~$A$ and~$B$.
We assume that these rates can change after each transition and depend on the current population counts.
We denote by~$\gamma_{k,A,B}$ the birth rate for each bacterium after the $k$\textsuperscript{th} transition, and by~$\delta_k$ the death rate for each bacterium after the $k$\textsuperscript{th} transition.

Denote by $P_k(A,B)$ the probability of species~$B$ being extinct before species~$A$ with indirect competition only, starting with the populations $(A,B)$ right after the $k$\textsuperscript{th} transition.
Ultimately we are interested in the case $k=0$, i.e., the probability $P_0(A,B)$.
Almost-sure consensus can be achieved only if one of the species gets extinct almost surely.
This requirement translates into a condition on the sequence of birth and death rates.
In a multiple-resource model, it is the case in particular.

\begin{lemma}\label{lem:indirect:probs}
If the species~$A$ and~$B$ are symmetric and get extinct almost surely, then
we have
$\displaystyle P_k(A,B) = \frac{A}{A+B}$
whenever $A+B \geq 1$.
\end{lemma}
\begin{proof}
The probabilities $P_k(A,B)$
are bounded between~$0$ and~$1$ and
satisfy the recurrence
\begin{equation}\label{eq:indirect:prob:recurrence}
\begin{split}
P_k(A,B)
& =
\frac{\gamma_{k,A,B}}{\gamma_{k,A,B}+\delta_{k,A,B}}
\left( \frac{A}{A+B} P_{k+1}(A+1,B)
+
\frac{B}{A+B} P_{k+1}(A,B+1) \right)
\\ & \quad +
\frac{\delta_{k,A,B}}{\gamma_{k,A,B}+\delta_{k,A,B}}
\left( \frac{A}{A+B} P_{k+1}(A-1,B)
+ \frac{B}{A+B} P_{k+1}(A,B-1) \right)
\end{split}
\end{equation}
for $A\geq1$ and $B\geq1$ with the boundary conditions $P_k(0,B)=0$ and $P_k(A,0)=1$.

It is straightforward to verify that $P_k(A,B)=\frac{A}{A+B}$ satisfies this recurrence:
\begin{equation}
\begin{split}
P_k(A,B)
& =
\frac{\gamma_{k,A,B}}{\gamma_{k,A,B}+\delta_{k,A,B}}
\left( \frac{A}{A+B} \cdot \frac{A+1}{A+B+1}
+
\frac{B}{A+B}  \cdot \frac{A}{A+B+1} \right)
\\ & \quad +
\frac{\delta_{k,A,B}}{\gamma_{k,A,B}+\delta_{k,A,B}}
\left( \frac{A}{A+B} \cdot \frac{A-1}{A+B-1}
+ \frac{B}{A+B} \cdot \frac{A}{A+B-1} \right)
\\
& =
\frac{\gamma_{k,A,B}}{\gamma_{k,A,B}+\delta_{k,A,B}}
\cdot \frac{A^2 + A + AB}{(A+B)(A+B+1)}
\\ & \quad +
\frac{\delta_{k,A,B}}{\gamma_{k,A,B}+\delta_{k,A,B}}
\cdot \frac{A^2 - A + AB}{(A+B)(A+B-1)}
\\
& =
\frac{\gamma_{k,A,B}}{\gamma_{k,A,B}+\delta_{k,A,B}}
\cdot \frac{A}{A+B}
+
\frac{\delta_{k,A,B}}{\gamma_{k,A,B}+\delta_{k,A,B}}
\cdot  \frac{A}{A+B}
=
 \frac{A}{A+B} \enspace .
\end{split}
\end{equation}

To prove uniqueness, let $P_k(A,B)$ and $\hat{P}_k(A,B)$ be two solutions of the recurrence that are bounded between~$0$ and~$1$, and set $\Delta_k(A,B) = P_k(A,B) - \hat{P}_k(A,B)$.
We will prove $\Delta_k(A,B)=0$ by showing $\lvert \Delta_k(A,B) \rvert \leq \varepsilon$ for all $\varepsilon > 0$.

Let $\varepsilon > 0$.
We write $\mathcal{T} = \{ A\!\!\nearrow, A\!\!\searrow, B\!\!\nearrow, B\!\!\searrow \}$ for the set of possible transitions, i.e., birth/death of~$A$ and birth/death of~$B$.
The differences $\Delta_k(A,B)$ satisfy the same recurrence
as $P_k(A,B)$, but with the boundary condition $\Delta_k(0,B)=\Delta_k(A,0)=0$.
We can rewrite the recurrence as
\begin{equation}\label{eq:delta:recurrence:single}
\begin{split}
\Delta_k(A,B)
=
\sum_{\tau\in\mathcal{T}} \Pr[ \tau_{k+1}=\tau \mid A(k)=A\,,\,B(k)=B ] \cdot \Delta_{k+1}(\tau(A,B))
\end{split}
\end{equation}
for all $A,B\geq 1$,
where we denoted the $(k+1)$\textsuperscript{th} transition by $\tau_{k+1}$ and used the transition~$\tau$ as an operator on the bacterial species counts to update them to their new values after the transition.
That is, we define $\tau(A,B) = (A+1,B)$ for the case $\tau = A\!\!\nearrow$ and analogously for the three other transitions.

Using the recurrence~\eqref{eq:delta:recurrence:single} multiple times leads to the formula
\begin{equation}\label{eq:delta:recurrence:multiple}
\begin{split}
\Delta_k(A,B)
=
\sum_{\substack{\sigma\in\mathcal{T}^\ell\\\sigma \text{ is live}}} \Pr\!\left[ (\tau_r)_{r=k+1}^{k+\ell} =\sigma \mid A(k)=A\,,\,B(k)=B \right] \cdot \Delta_{k+\ell}(\sigma(A,B))
\end{split}
\end{equation}
where we call a sequence $\sigma \in \mathcal{T}^\ell$ \emph{live} if neither species~$A$ nor~$B$ is extinct after any of the transitions in~$\sigma$ when starting with the populations $(A,B)$ right after the $k$\textsuperscript{th} transition.
All terms of the sum that do not correspond to live sequences are zero because of the boundary condition for~$\Delta_{k+\ell}$.

Because of the almost-sure extinction hypothesis we have:
\begin{equation}
\lim_{\ell\to\infty}
\Pr\!\left[ (\tau_r)_{r=k+1}^{k+\ell} \text{ is live} \right] 
= 0 \enspace .
\end{equation}
There hence exists an~$\ell\geq 1$ such that the probability of $(\tau_r)_{r=k+1}^{k+\ell}$ being live is less than or equal to~$\varepsilon$.
Since the bounds on the solutions guarantee $\lvert \Delta_{k+\ell}(\sigma(A,B))\rvert \leq 1$, we get:
\begin{equation}
\begin{split}
\lvert \Delta_k(A,B) \rvert
& \leq
\sum_{\substack{\sigma\in\mathcal{T}^\ell\\\sigma \text{ is live}}} \Pr\!\left[ (\tau_r)_{r=k+1}^{k+\ell} =\sigma \mid A(k)=A\,,\,B(k)=B \right] \cdot \big\lvert \Delta_{k+\ell}(\sigma(A,B)) \big\rvert
\\ & \leq 
\Pr\!\left[ (\tau_r)_{r=k+1}^{k+\ell} \text{ is live} \right]
\leq 
\varepsilon \enspace .
\end{split}
\end{equation}

We can thus conclude $\Delta_k(A,B)=0$, which proves that $P_k(A,B)=\frac{A}{A+B}$ is the unique solution of recurrence~\eqref{eq:indirect:prob:recurrence} that satisfies the boundary conditions and is bounded between~$0$ and~$1$.
\end{proof}

To generalize to non-symmetric species, we denote by~$\gamma^{(X)}_{k,A,B}$ and~$\delta^{(X)}_{k,A,B}$ the birth and death rates of species $X\in\{A,B\}$ after the $k$\textsuperscript{th} transition in configuration~$(A,B)$, respectively.
We say that species~$X$ \emph{dominates} species~$Y$ if 
$\gamma^{(X)}_{k,A,B} \geq \gamma^{(Y)}_{k,A,B}$
and
$\delta^{(X)}_{k,A,B} \leq \delta^{(Y)}_{k,A,B}$.

\begin{theorem}
If the species~$A$ and~$B$ are symmetric or if one dominates the other,
indirect competition cannot guarantee that majority consensus is achieved with a probability larger than the relative initial population of the majority species.
\end{theorem}
\begin{proof}
The lack of resource in-flow guarantees almost-sure extinction of both species.
If the species are symmetric, then the claim hence follows from Lemma~\ref{lem:indirect:probs}.
If one species dominates the other, starting the system in a configuration in which the dominating species is in the minority, the probability of the dominated, majority species winning is upper-bounded by the probability of it winning in a system with symmetric species.
\end{proof}

\section{Simulation Case Study: A Synthetic Plasmid Conjugation System}
\label{sec:simulations}

We complement our analytical results with simulations to validate that the
asymptotics shown in the previous sections can already be observed in realistic biological settings.
As a proof-of-principle study, we chose a hypothetical synthetic, engineered system for two directly competing bacteria $A$ and $B$.
The competition is assumed to be engineered based on conjugation, a process where a small circular DNA in the form of a plasmid is exchanged between two bacteria upon physical contact and subsequent plasmid transfer.

\subsection{An Engineered Mutual Annihilation Protocol}

We modeled a culture of two bacterial types, $A$ and $B$, 
that grow in a closed system of volume~$1\,\mu \text{L}$. 
Both bacterial types have identical growth behavior, feeding on two resources
$R_1$ and $R_2$ that model components of the medium with different nutrient efficiencies.
Such behavior has been observed for \emph{E. coli} grown in lab conditions \cite{sezonov2007escherichia}.
The duplication reactions for cells $C \in \{A,B\}$ are thus
\begin{equation}
C+R_1 \xrightarrow{\makebox[0.6cm]{$\gamma_1$}} 2C \qquad
C+R_2 \xrightarrow{\makebox[0.6cm]{$\gamma_2$}} 2C\enspace,
\end{equation}
where we set the single-cell growth rates to 
\[
\gamma_1(R_1) = R_1/R_1(0) \cdot 1/20\,\text{min}^{-1} \quad \textrm{ and } \quad \gamma_2(R_2) = R_2/R_2(0) \cdot 1/20 \cdot 0.08\,\text{min}^{-1}\enspace.
\]
This corresponds to an initial expected duplication time of $20\,\text{min}$
  with the help of resource $R_1$, and only $8\%$ of that duplication rate using resource~$R_2$.
As initial resource concentrations, we chose $[R_1] = 2\cdot 10^{6} \,\text{mL}^{-1}$ and $[R_2] = 10^{8} \,\text{mL}^{-1}$, resulting in a carrying capacity of about $10^8$ bacteria per $\text{mL}$.
  Observe that the growth rates are bounded, as required by our model, since resources are
  not regenerated and there is no resource inflow.

Further, bacteria die with an individual death rate that was
set to $\delta = 10^{-4}\,\text{min}^{-1}$, which is on the same order as $0.43\,\text{d}^{-1}$, the rate measured~\cite{schink2019death} for \emph{E. coli}.
We modeled two types of systems: one following direct competition, and for control, one without direct competition.
For the direct competition, we assumed that each of the bacteria carries a
  respective plasmid that,
  if introduced via conjugation into the other bacterial type upon an interaction, leads to its death.
For a more refined setting, instead of immediate death, we consider a short-lived intermediate bacterial type $AB$ with increased death rate for bacteria that carry both plasmids.
We thus have
\begin{equation}
A+B \xrightarrow{\makebox[0.6cm]{$\alpha$}} A + AB \qquad
A+B \xrightarrow{\makebox[0.6cm]{$\alpha$}} AB + B \qquad
AB \xrightarrow{\makebox[0.6cm]{$\alpha'$}} \emptyset
\enspace,
\end{equation}
where we chose $\alpha = 5 \cdot 10^{-10} \,\text{mL}^{-1}\text{min}^{-1}$ in accordance with measurements of conjugating \emph{E. coli}~\cite{wan2011measuring} and an increased death rate of $\alpha' = 10^{-1}\,\text{min}^{-1}$.
Table~\ref{tab:parameters} summarizes the parameterization of the model.

\begin{table}[]
\centering
\begin{tabular}{@{}l@{\quad}l@{\quad}p{4.5cm}@{}}
\toprule
Parameter & Value & Note \\ 
\midrule
Cell death rate const.\ &
$\delta=10^{-4}\,\text{min}^{-1} $ &
order as in \cite{schink2019death} \\
Cell growth rate for $R_1$ &
$\gamma_1(R_1) = R_1/R_1(0) \cdot 1/20\,\text{min}^{-1}$ &
$20\,\text{min}$ duplication time\\
Cell growth rate for $R_2$ & 
$\gamma_2(R_2) = R_2/R_2(0) \cdot 1/20 \cdot 0.08\,\text{min}^{-1}$ &
reduced nutrition efficiency\\ 
Conjugation rate const.\ &
$\alpha = 5 \cdot 10^{-10} \,\text{mL}^{-1}\text{min}^{-1}$ &
from \cite{wan2011measuring}\\
Increased death rate const.\ &
$\alpha' = 10^{-1} \,\text{min}^{-1}$ &
death after $10\,\text{min}$\\
\bottomrule
\end{tabular}
\caption{Parameters of the simulation model.\label{tab:parameters}}

\end{table}

\subsection{Simulation Results}

We ran stochastic simulations of two competing bacteria populations, $A$ and $B$, with initial concentrations of $[A] + [B] = 3\cdot 10^5 \mu\text{L}^{-1}$. 
We compared the performance of the mutual annihilation protocol to the case of resource-consumer dynamics without direct competition. 

\begin{figure}[t]
    \centering
    \begin{minipage}[t]{.4\textwidth}
        \centering
          \includegraphics[width=5.5cm]{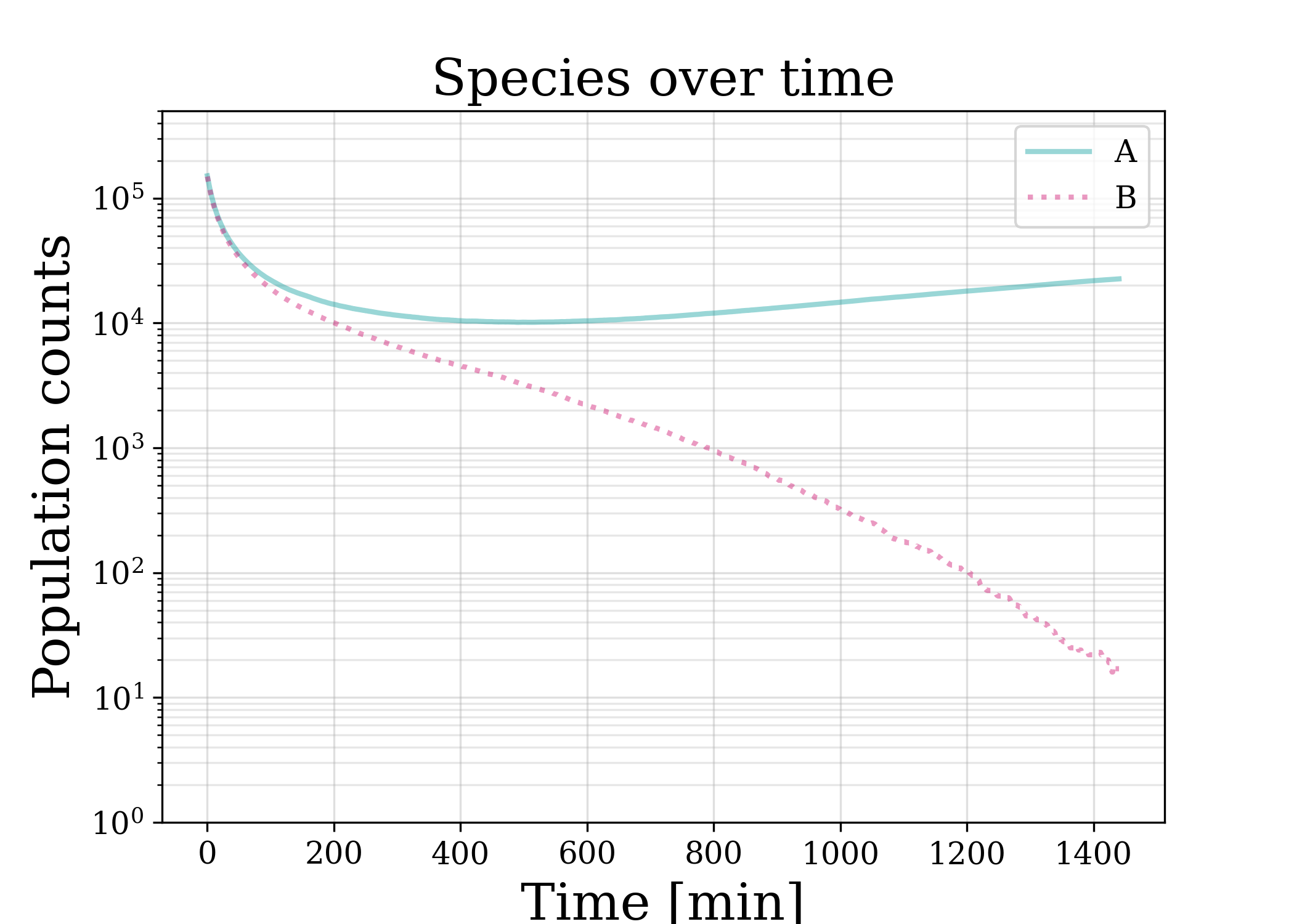}
          \caption{Stochastic simulation with initial population counts $A = 151000$ and
          $B = 149000$ over $1$\,day. Population counts are
          per $\mu$L and plotted on a logarithmic scale.}
          \label{fig:run}
    \end{minipage}\hfill%
    \begin{minipage}[t]{.25\textwidth}
        \centering
        \includegraphics[width=3.8cm]{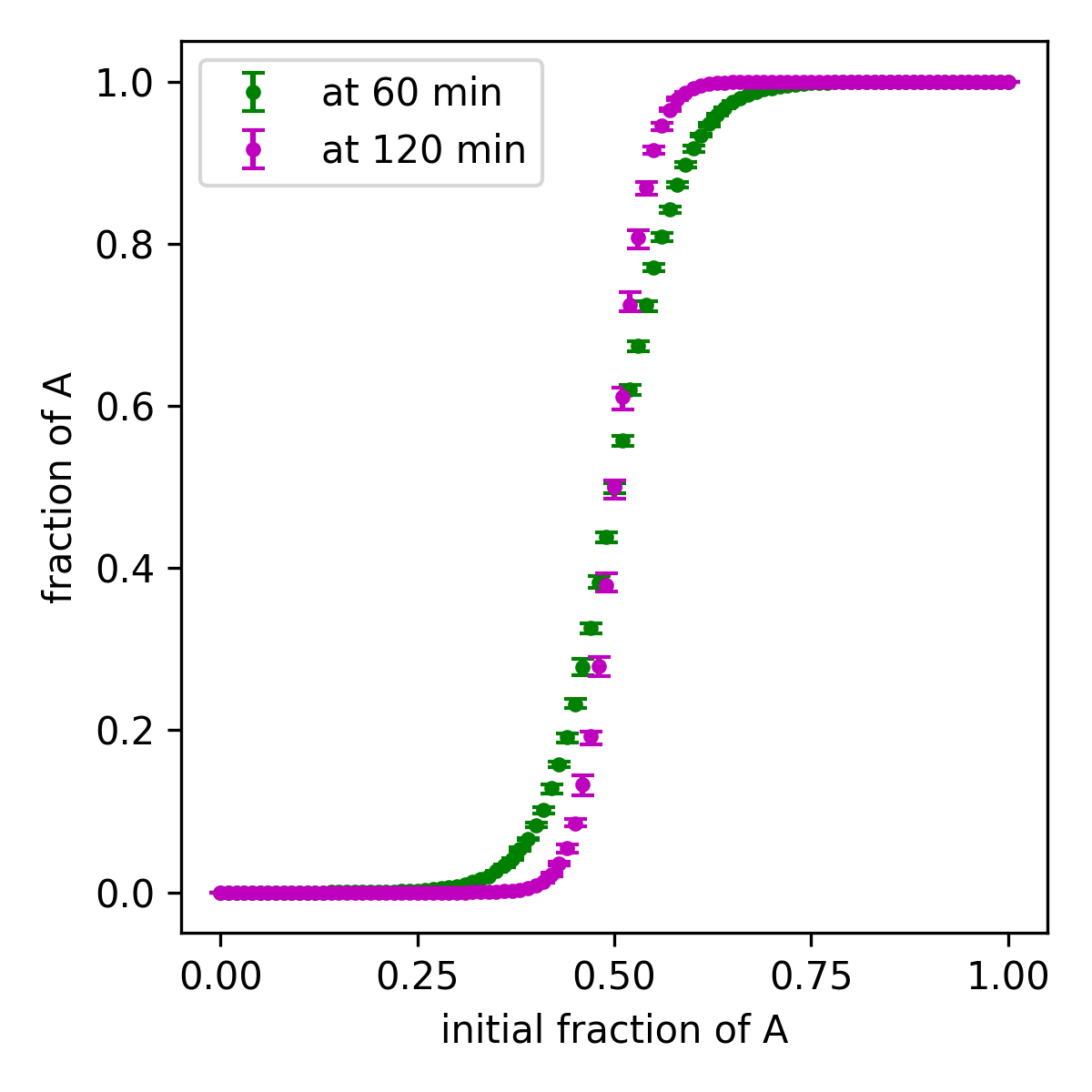}
        \caption{Fraction of $A$ in the bacterial population after $60$\,min and $120$\,min. $N=10$ simulations per initial fraction. Error bars indicate maximum and minimum, and markers indicate average fractions.}
        \label{fig:s_curve}
    \end{minipage}\hfill%
    \begin{minipage}[t]{0.25\textwidth}
        \centering
        \includegraphics[width=4cm]{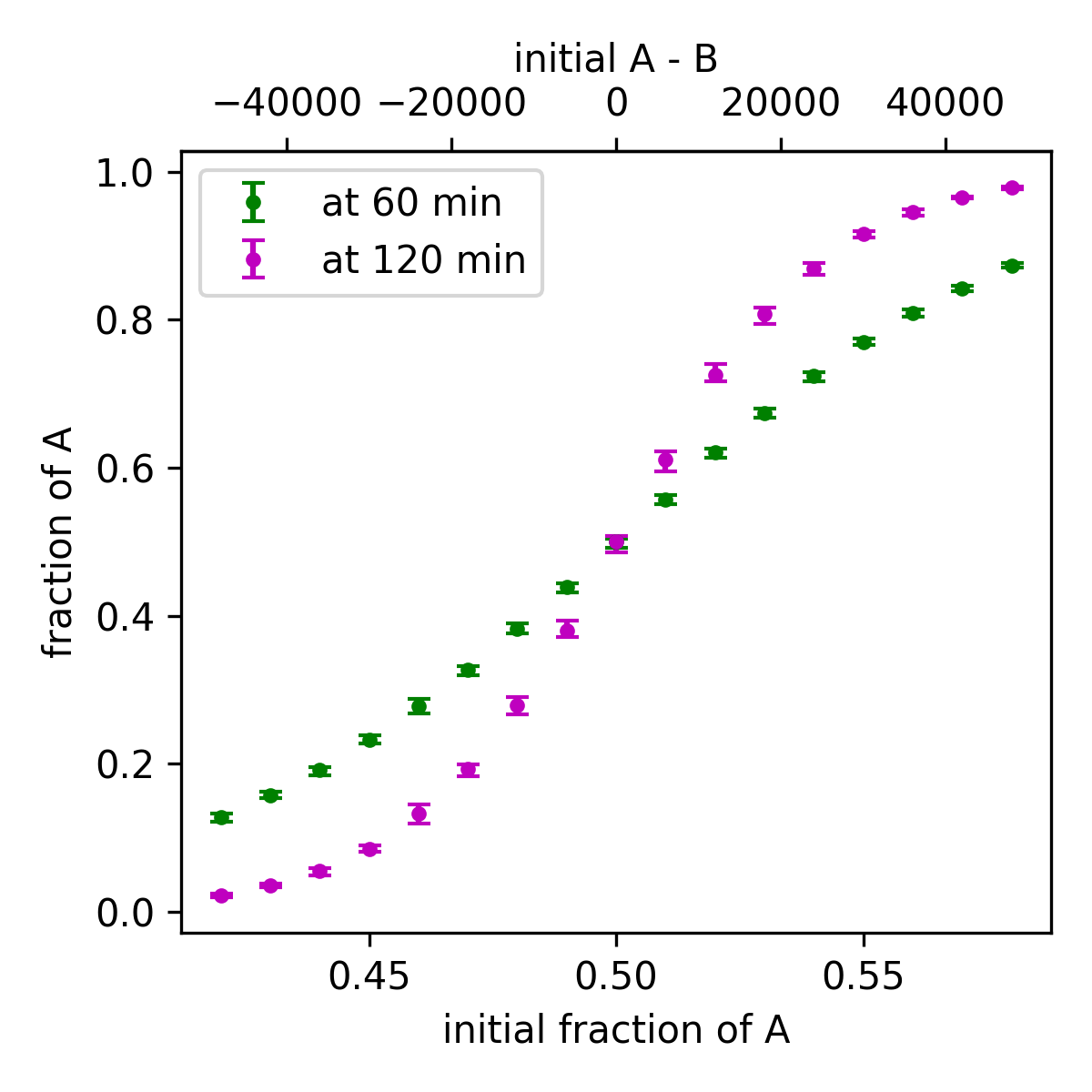}
        \caption{Zoomed version with initial population difference on the top abscissa.}
        \label{fig:s_curve_zoom}
    \end{minipage}
\end{figure}

\paragraph{Direct competition as an amplifier}
Figure~\ref{fig:run} shows a single stochastic trajectory during the first $1400$\,min of simulated time with a total initial population of $A+B = 3\cdot 10^5$.
Here, the initial population counts of $A$ and $B$ have been set to
  differ by $2000$ with $A$ being the majority.
  We can clearly see that the initial difference between the two population sizes is amplified over time: after $1400$\,min (about one day), the count of the minority species has decreased by three orders of magnitude.
  
Figure~\ref{fig:s_curve} shows the prevalence of type $A$ after $60$\,min and $120$\,min as a function of the initial fraction of type $A$ individuals, varying from $0$ to $1$.
Observe the steep s-shaped behavior that is typical for a large amplification away
  from the midpoint of equal concentrations.
Figure~\ref{fig:s_curve_zoom} zooms into the middle of the s-shaped curve, with the top abscissa showing the difference $A-B$.
  
\paragraph{Comparison with resource-consumer dynamics}
The dynamics of the mutual annihilation protocol show that direct competition quickly amplifies the differences between the two populations. We also compared this scenario to a setting in the absence of direct competition. To this end, we set the conjugation rate parameter $\alpha=0$.
In this case, competition is mediated only by consumption of shared resources.
  
\begin{figure}[t]
    \centering
    \begin{minipage}[t]{.4\textwidth}
        \centering
          \includegraphics[width=5.5cm]{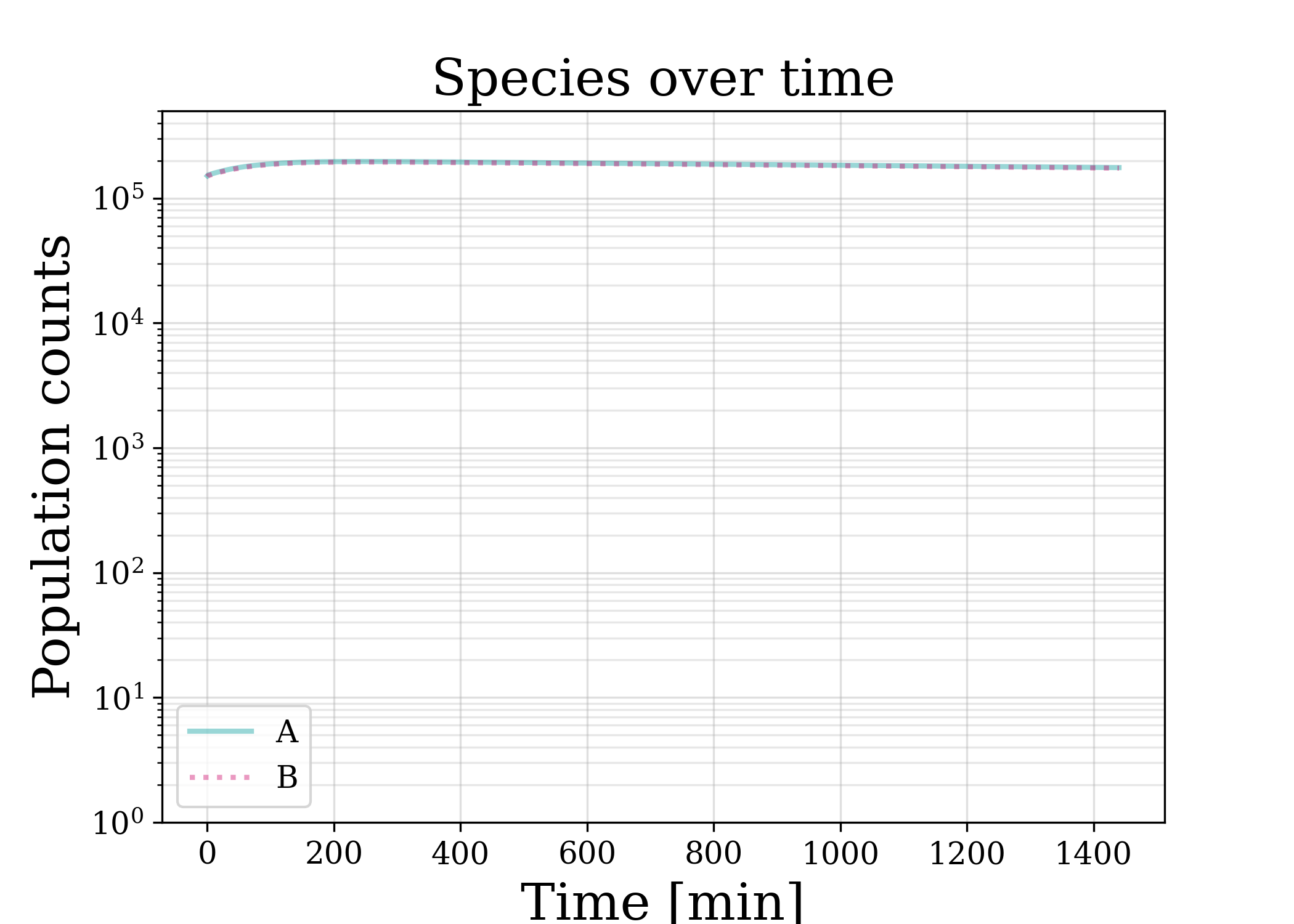}
          \caption{Stochastic simulation as in Figure~\ref{fig:run}, but with $\alpha = 0$ and shown over $1$\,day.}
          \label{fig:run_noab}
    \end{minipage}\hfill%
    \begin{minipage}[t]{.25\textwidth}
        \centering
        \includegraphics[width=3.8cm]{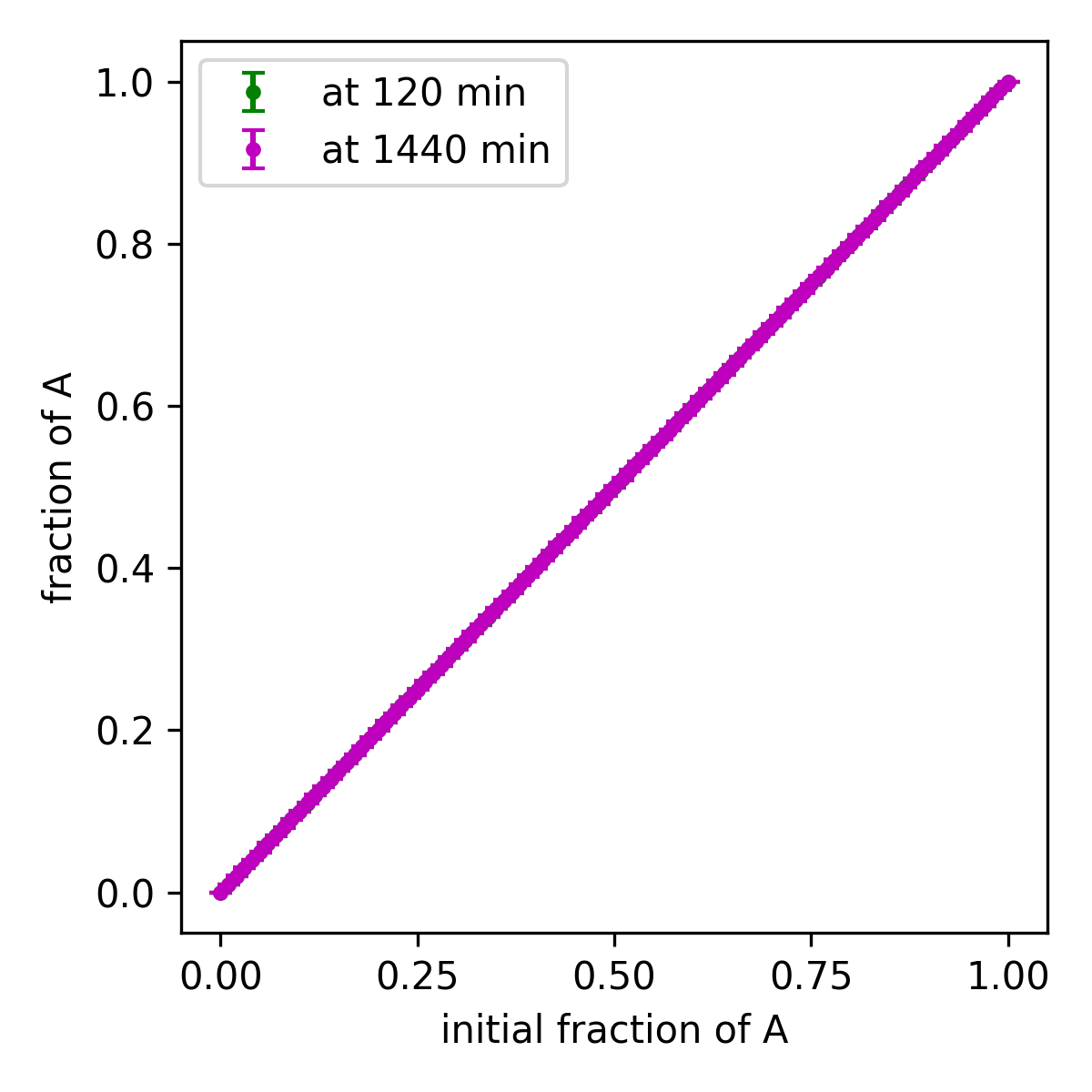}
        \caption{Same setting as in Figure~\ref{fig:s_curve}, but with $\alpha = 0$ and snapshot after one day instead of $60$\,min.}
        \label{fig:s_curve_noab}
    \end{minipage}\hfill%
    \begin{minipage}[t]{0.25\textwidth}
        \centering
        \includegraphics[width=4cm]{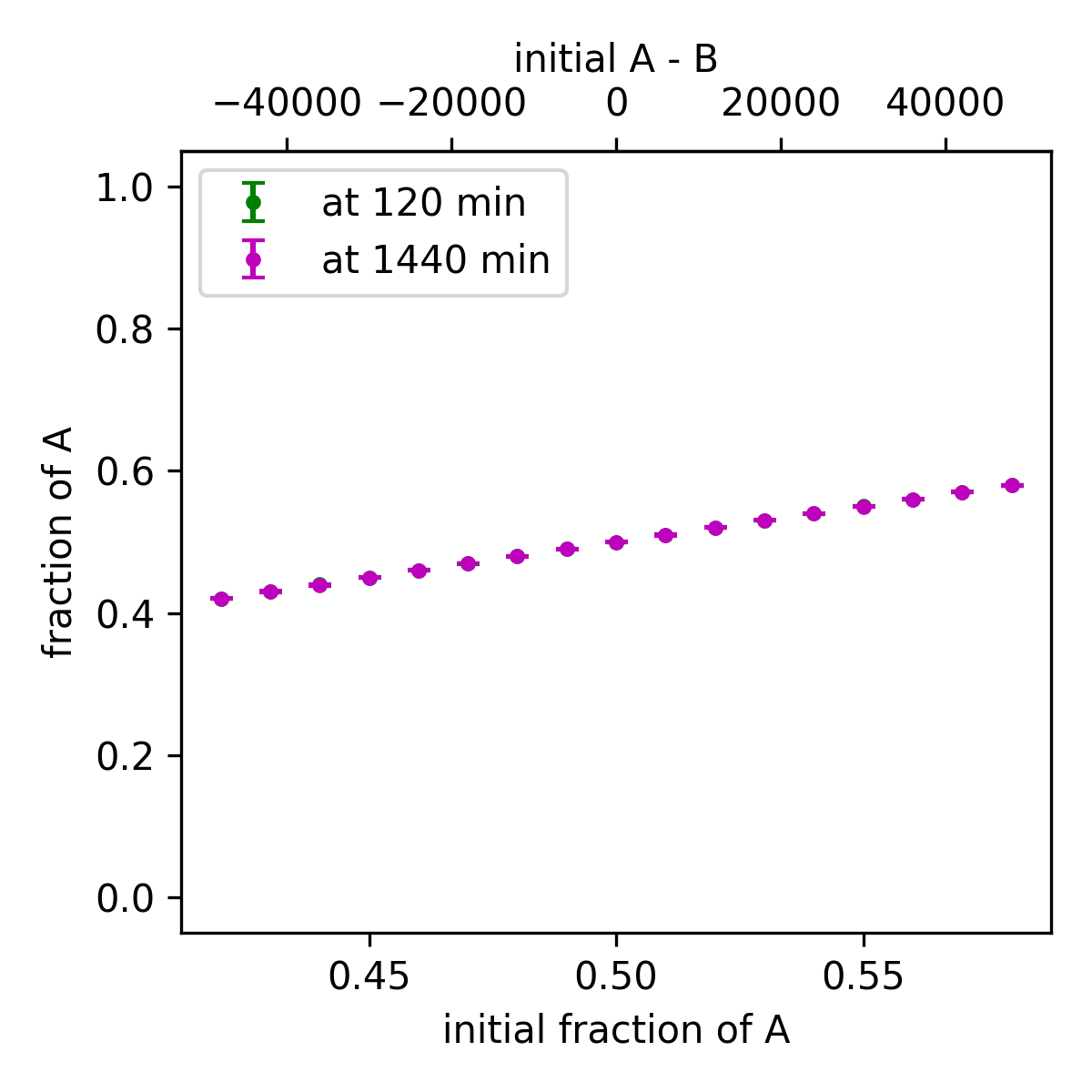}
        \caption{Zoomed version corresponding to Figure~\ref{fig:s_curve_zoom}, but with $\alpha = 0$ and snapshot after one day instead of $60$\,min.}
        \label{fig:s_curve_zoom_noab}
    \end{minipage}
\end{figure}

Figures~\ref{fig:s_curve_noab} and~\ref{fig:s_curve_zoom_noab} show the obtained results after $60$\,min of simulated time. We can see that compared to the case of interference competition, there is little to no amplification of the differences under exploitative competition during the first 60\,min of simulated time. Moreover, this holds even after 1 day, as shown by Figure~\ref{fig:run_noab}.

\section{Conclusion}

We have investigated distributed consensus in microbial systems subject to stochastic population dynamics and competitive interactions. Our main result establishes that direct interference competition, modeled through the mutual annihilation protocol, efficiently solves majority consensus with high probability when the initial gap is $\Omega(\sqrt{n\log n})$, where $n$ is the total population size. By extending the coupling technique of Cho et al.~\cite{CFHKNS21:growth} to a new model, we showed that majority consensus is reached in~$O(n)$ steps with high probability.

In contrast, indirect competition through shared resources alone cannot reliably achieve majority consensus. In symmetric systems, the probability of reaching majority consensus equals merely the initial relative population of the majority species.

Our simulations with realistic biological parameters demonstrate that these theoretical results manifest at practical time scales. Using engineered \emph{E.\ coli} with plasmid conjugation-based competition, we observed rapid amplification of initial differences and majority consensus within one day, while systems without direct competition showed negligible amplification over the same period.

This work opens several directions for future research, including analyzing asymmetric species, extending to open environments with resource inflow, investigating more complex distributed tasks beyond binary consensus, and experimental implementation in synthetic bacterial systems.

\section*{Ethical Approval}

Not applicable.

\section*{Competing Interests}

The authors declare no competing interests.

\section*{Authors' Contributions}

M.F., T.N., and J.R. conceived the study, performed the mathematical analysis, and wrote the paper. V.A., M.F., B.M., and T.N. performed the simulations. All authors reviewed the manuscript.

\section*{Funding}
This work was partially supported by the ANR project DREAMY (grant ANR-21-CE48-0003) and the CNRS projects ABIDE and BACON.

\section*{Availability of Data and Materials}

No data associated with the manuscript.

\bibliography{references}

\end{document}